\documentstyle[preprint,aps,prb]{revtex}
\begin{document}
\draft
\title{Structure and Magnetism of Fe/Si Multilayers Grown by
Ion-Beam Sputtering}
\author{A. Chaiken\cite{Aemail}, R.P. Michel\cite{Remail} and M.A.
Wall\cite{Memail}}
\address{Materials Science and Technology Division\\Lawrence
Livermore National Lab\\Livermore, CA 94551}
\date{\today}

\maketitle

\begin{abstract}
Ion-beam sputtering has been used to prepare Fe/Si multilayers on a
variety of substrates and over a wide range of temperatures.
Small-angle x-ray diffraction and transmission electron microscopy
experiments show that the layers are heavily intermixed although a
composition gradient is maintained.  When the spacer layer is an
amorphous iron silicide, the magnetic properties of the multilayers
are similar to those of bulk Fe.  When the spacer layer is a
crystalline silicide with the B2 or DO$_3$ structure, the multilayers
show antiferromagnetic interlayer coupling like that observed in
ferromagnet/paramagnet multilayers such as Fe/Cr and Co/Cu.  Depending
on the substrate type and the growth temperature, the multilayers grow
in either the (011) or (001) texture.  The occurrence of the
antiferromagnetic interlayer coupling is dependent on the
crystallinity of the iron and iron silicide layers, but does not seem
to be strongly affected by the perfection of the layering or the
orientation of the film.  Since the B2- and DO$_3$-structure Fe$\rm
_x$Si$\rm _{1-x}$ compounds are known to be metallic,
antiferromagnetic interlayer coupling in Fe/Si multilayers probably
originates from the same quantum-well and Fermi surface effects as in
Fe/Cr and Co/Cu multilayers.

\end{abstract}
\pacs{75.50.Bb,61.10.-i,68.65.+g,81.15.Cd,75.30.E}
\clearpage

\section{Introduction}

Multilayer films formed from transition metals and semiconductors have
long been studied because of their unusual superconducting
properties\cite{ruggiero} and because of possible application as x-ray
optical elements.\cite{barbee85} Many unusual phenomena have been
produced, ranging from the observation of dimensional crossover in
weakly coupled superconducting Nb layers in Nb/Ge
multilayers\cite{ruggiero} to the occurrence of bcc Ge in short-period
Mo/Ge multilayers.\cite{wilson} Unusual magnetic properties have
recently been observed in Fe/Si multilayers by workers at
ETH\cite{toscano} and Argonne.\cite{fullerton} A large
antiferromagnetic (AF) interlayer coupling in these multilayers
manifests itself in hysteresis loops as a high saturation field and a
low remanent magnetization.  Similar magnetization curves are
associated with large interlayer coupling in metal/metal multilayers
like Fe/Cr and Co/Cu.\cite{demokritov,mosca} Much consideration has
been given to whether the coupling in the Fe/Si system has the same
origin as in the metal/metal multilayers.\cite{bruno,singh} Therefore
the question of whether the spacer layer in the Fe/Si multilayers is a
metal or semiconductor is of particular interest.

Previous work on Nb/Si,\cite{fullerton2} Co/Si,\cite{miura}
Ni/Si,\cite{wang} and Mo/Si\cite{stearns} multilayers have shown that
there is a strong tendency towards compound formation at the
metal/silicon interface.  In general these multilayers consist of
polycrystalline metal layers separated by an amorphous silicon layer
which is bounded on either side by a layer of intermixed material.
The intermixed silicide layers in these films were amorphous unless
they were annealed at several hundred
$^{\circ}$C.\cite{miura,holloway} These previously studied multilayers
were therefore likely in their as-grown state to have
metal/semiconductor character because of the presence of the amorphous
silicon layer.

In order to investigate the character of the spacer layer in the Fe/Si
multilayer system, we have grown a large number of films with
different substrate temperatures, substrate types, and layer
thicknesses.  When the Si spacer layer thickness is greater than about
20\AA, we find that the metal layers are crystalline but that the
spacer layers are amorphous, similar to the situation in other
transition metal/silicon systems.  When the Si spacer layer thickness
is less than about 20\AA thick, the iron silicide spacer layer forms a
crystalline silicide with either the B2 or DO$_3$ structure.  The B2
structure consists of two interpenetrating simple cubic sublattices
and is identical to the CsCl structure for a 1:1 ratio of Fe and
Si,\cite{vonkanel} while the DO$_3$ structure is an fcc lattice with
two inequivalent Fe sites.\cite{kudrnovsky} Extensive growth
experiments, described below, suggest that crystallinity of the spacer
layer is crucial for occurrence of the antiferromagnetic interlayer
coupling, in keeping with previous suggestions.\cite{fullerton} Since
both the B2 and DO$_3$ phases are metallic,\cite{vonkanel,kudrnovsky}
the fact that crystallinity is required for antiferromagnetic coupling
suggests that the coupling in Fe/Si has a common origin with that
observed in metal/metal multilayers.

\section{Experimental Methods}

The Fe/Si multilayers are  grown in the ion-beam sputtering
(IBS) chamber whose layout is shown schematically in
Figure~\ref{chamber}.  The system base pressure is typically about
2$\times$10$^{-8}$ torr. The ion gun is a 3 cm Kauffman source with
focusing optics.\cite{commonwealth} The energy of the ions leaving the
gun can be modulated by raising and lowering the voltage on the
acceleration grids, creating in effect an electrical shutter.  The Ar
ions are incident on the sputter target at 1000V at an angle of about
45$^{\circ}$.  The Ar pressure is maintained in the
2-3$\times$10$^{-4}$ range by a flow-controller coupled to a
capacitance manometer.\cite{MKS} Four 3" diameter sputter targets are
mounted on a tray which can be rotated by a stepper motor.\cite{MDC}
Layer thickness is monitored by a quartz-crystal oscillator which is
placed in close proximity to the substrates.  The substrates are about
25 cm above the targets, clamped to a copper tray.  The temperature of
the tray is monitored by a thermocouple and can be varied between
-150$^{\circ}$C and +200$^{\circ}$C.\cite{rsipaper} Three films are
grown per chamber pumpdown.

The thickness monitor, the controller for the stepper motor and the
ion-beam power supply are all interfaced to a personal computer which
has been programmed using the ASYST instrument control
package.\cite{keithley} When the system is depositing a multilayer,
the computer sends the material parameters to the thickness monitor,
rotates the stepper motor to its new orientation, and turns the ion
gun on.  When the desired thickness is reached, the thickness monitor
turns the ion-gun off and prompts the computer for the next layer.
The basic design of the system is similar to one previously described
by Kingon {\it et al.}\cite{kingon}

The substrates for multilayers growth include glass coverslips,
oxidized silicon wafers, MgO (001) and Al$_2$O$_3$(0\={2}11).  The
first two substrates, which are used for growth of polycrystalline
films, were rinsed in solvents before loading into the vacuum chamber.
The second two, which are used for epitaxial growth, are cleaned
according to a recipe reported by Farrow and coworkers.\cite{farrow}
The typical deposition rate for Fe is 0.2\AA/s while that for Si is
about 0.3\AA/s.  All films are capped with a 200\AA\ Ge oxidation
barrier.  The magnetic and structural properties of the films are
stable for at least one year.  Ge is used for capping instead of Si in
order to prevent interference with element-specific soft x-ray
fluorescence measurements, which will be reported
elsewhere.\cite{carlisle}

X-ray diffraction characterization has been performed using a 18kW
rotating anode system outfitted with a graphite monochromator.  All
spectra are taken using the Cu K$_{\alpha}$ wavelength.  Conventional
high-resolution electron microscopy and electron diffraction have been
performed in order to characterize the microstructure of the
as-deposited films in cross-section.  Magnetization curves are
obtained using a vibrating sample magnetometer.  All the data shown
here were taken at room temperature.

\section{Results}

Overall the magnetic properties of the Fe/Si multilayers made by IBS
are similar to those made previously by magnetron
sputtering.\cite{fullerton,foiles} Definitive confirmation of AF
interlayer coupling in our multilayers has been obtained by polarized
neutron reflectivity measurements.\cite{ankner} For some unknown
reason the magnetic properties of our multilayers are closer to those
of the magnetron-sputtered multilayers than those reported on in a
previous study using IBS, where much lower saturation fields were
observed.\cite{inomata} The differences between the previous IBS-grown
films and ours may be related to the lower ion-beam voltage used by
Inomata {et al.}\cite{inomata} Comparisons on the basis of layer
thickness are made here only between films grown during the same
deposition run in order to insure that the relative layer thicknesses
are meaningful.  Films with similar layer thicknesses have been grown
many times to establish reproducibility of the observed trends.

\subsection{Layer-Thickness Dependence of Properties}
\label{thicksec}

Forty- and fifty-repeat multilayers have been grown with t$\rm_{Fe}$ =
14, 20, 30, 40, and 50\AA\ and t$\rm _{Si}$ = 14 and 20\AA.
Magnetization curves for 50-repeat (Fe30\AA/Si20\AA) and
(Fe30\AA/Si14\AA) multilayers grown on glass at nominal RT (about
+60$^{\circ}$C) are shown in Figure~\ref{sithick}.  On the y-axis of
this plot is the magnetic moment of the multilayer normalized to the
moment of an equivalent volume of bulk Fe.  The magnetization curve of
the (30/20) multilayer looks much like that of an Fe film, while the
magnetization curve of the (30/14) multilayer shows the high
saturation field and low remanence which characterize AF interlayer
coupling.  At its saturation field the magnetization of the (30/14)
multilayer is about the same as for the (30/20) multilayer.  Both of
these films have moments only about half as large as an equivalent
volume of bulk Fe.  Our observation of AF coupling for Si thicknesses
between 10 and 20\AA\ and the disappearance of coupling for Si thicker
than 20\AA\ confirm previous observations on magnetron-sputtered
films.\cite{fullerton}

X-ray diffraction spectra for these multilayers are shown in
Figures~\ref{sithickxrdlow} and ~\ref{sithickxrdhi}.
Figure~\ref{sithickxrdlow} shows the small-angle x-ray scattering
(SAXS) data with peaks at angles

\begin{equation}
n^2 \lambda^2 \, = \, 4 \Lambda^2 sin^2 \theta \; + \; 2\delta
\label{lowbragg}
\end{equation}

\noindent where $\lambda$ is the x-ray wavelength, $\rm \Lambda \:
\equiv \: t_{Fe} \: + \: t_{Si}$ is the multilayer bilayer period and
$\delta$ is the index of refraction for x-rays.\cite{fullerton3} This
grazing incidence data gives information about the quality of the
multilayer interfaces.  Figure~\ref{sithickxrdlow} shows four
low-angle peaks for both films, indicating a reasonably strong
composition modulation along the growth direction.  (The higher
frequency oscillations between 1$^{\circ}$ and 4$^{\circ}$ are
finite-thickness fringes from the Ge cap layer.)  The most notable
difference between the two spectra is that the multilayer peaks are
broader for the AF-coupled t$\rm _{Si}$ = 14\AA\ film, indicating more
fluctuations in bilayer period and probably more interface roughness.
Using the spacing between peak positions to eliminate the unknown
$\delta$ from Eqn.~\ref{lowbragg} gives values of the bilayer period
$\Lambda$ for the two films.  For the multilayer with nominal layering
of (Fe30\AA/Si20\AA)x50, the derived value for $\Lambda$ is (41.82
$\pm$ 0.07)\AA, while for the (Fe30\AA/Si14\AA)x50 film $\Lambda$ =
(38.10 $\pm$ 0.04)\AA.  $\Lambda$ is 8.2\AA\ shorter than the nominal
value for the t$\rm _{Si}$ = 20\AA\ film, and 5.9\AA\ shorter than
nominal for the t$\rm _{Si}$ = 14\AA\ film.  Although some of the
discrepancy between the nominal and observed bilayer period may be due
to calibration inaccuracies, most is undoubtedly due to intermixing of
the Fe and Si layers, in keeping with observations in the other
metal/Si multilayers.\cite{fullerton2,holloway} Throughout this paper
we will continue for convenience to refer to the films in terms of
their nominal layer thicknesses.

Comparison of the magnetization data to the x-ray data can give some
further insight into the question of intermixing.  Because of the
presumed interdiffusion of the Fe and Si layers, the magnetic moment
of the Fe layers is also reduced from the nominal value.  The missing
magnetic moment can be expressed as an equivalent thickness of Fe.
Figure~\ref{missmom} shows a plot of missing moment in units of \AA
ngstroms of Fe versus missing bilayer period determined from
multilayer peak positions in SAXS for films grown at room temperature
(RT).  The plot shows that while the diffusion-induced reduction in
bilayer period varies between 1 and 8\AA, the missing Fe moment per
bilayer (for both interfaces) is consistently between 10 and 12\AA.
The one outlier in Figure~\ref{missmom} is for a film which had t$\rm
_{Fe}$ = 20\AA, the thinnest Fe for which we have ever observed
interlayer coupling. Other groups have previously observed a moment
reduction of 12-14\AA\ per bilayer in polarized neutron reflectivity
measurements on uncoupled Fe/Si multilayers with thick Si
layers.\cite{dufour,ankner2}

The disparity between the magnetic moment reduction and the bilayer
period reduction numbers may at first appear to be puzzling.  This
disparity occurs because the moment and bilayer period are affected by
different aspects of the structure.  In calculating the moment
reduction in \AA\ the assumption has been made that the Fe layer has
the magnetization of bulk Fe.  This is equivalent to assuming that
there is no Si in the Fe layer, which is undoubtedly false.  In
calculating the missing bilayer period, the assumption has also been
made that the spacer layer is pure Si, also clearly false.  The fact
that the missing magnetic moment is almost constant irrespective of
the reduction in bilayer period suggests that the spacer layer is
non-magnetic independent of Si thickness.  The lack of variation of
the missing moment is then explained by the diffusion of a constant
number of iron atoms into the silicon layer, irrespective of its
thickness.  The wide variation of the measured bilayer period is most
likely related to the varying orientation and crystallinity of the
spacer layer, neither of which affects the magnetic moment if the
spacer itself is non-magnetic.

Figure~\ref{sithickxrdhi} shows the high-angle x-ray spectra where
peak positions give information about the orientation and
crystallinity of the films.  The intense peak near 70$^{\circ}$ in
this plot is due to the Si substrate.  Included are data for an
(Fe40\AA/Si14\AA)x40 antiferromagnetically coupled multilayer and for
an (Fe30\AA/Si20\AA)x40 uncoupled multilayer, both grown on oxidized
Si(001) at RT.  The peaks for the (40/14) film are narrower than for
the (30/20).  The Scherrer formula gives 78\AA\ or about two bilayer
periods for the coherence length of the (40/14) film and 34\AA\ or
about one bilayer period for the coherence length of the (30/20) film.
Coherence lengths in IBS-sputtered antiferromagnetically coupled films
are often as long as 200\AA.  Fullerton {\it et al.}  have inferred
that the spacer layer in thin-Si Fe/Si multilayers must be crystalline
based on their observation of coherence lengths longer than a bilayer
period.\cite{fullerton} In keeping with its superior crystallinity,
the (40/14) multilayer has one superlattice satellite on the low-angle
side of the Fe(002) peak.  Typically only one satellite on the
low-angle side of the Fe (011) or (002) x-ray peak is observed for
polycrystalline multilayers grown on glass, in agreement with
observations by Foiles {et al.}\cite{foiles2}

The thin-Si multilayers which have AF coupling usually show a mixed
[001] and [011] orientation when grown on glass substrates at RT.
Occasionally t$\rm _{Si}$ = 14\AA\ films with a pure (011) orientation
are obtained at RT.  The variation in texture may be due to changes in
film stress under slightly different deposition conditions.  Stress
induced during deposition has been postulated to explain the mixed Mo
texture found in Mo/Ge multilayers.\cite{wilson} In contrast to the
thin-Si case, the thicker-Si Fe/Si multilayers which do not show
interlayer coupling always have a pure (011) texture.  Since the (011)
plane is close-packed for the bcc crystal structure, one would expect
the (011) orientation to be energetically favored for the Fe in a
multilayer with amorphous Si.  Films grown at nominal RT on glass or
oxidized Si substrates typically had rocking curves about 10$^{\circ}$
wide indicating a moderate amount of orientation.

Transmission electron microscopy (TEM) has been used to further
investigate the morphology of the films.  TEM cross-sectional images
of an (Fe30\AA/Si20\AA)x50 multilayer and an (Fe40\AA/Si14\AA)x50
multilayer grown during the same deposition run are shown in
Figures~\ref{TEMlowres}a and \ref{TEMlowres}b, respectively.  The most
salient features of the (30/20) multilayer are the long lateral
continuity of the layers and the smoothness of the interfaces.  Since
there is no interlayer coherence in the (30/20) film, the crystalline
grains have a high aspect ratio.  The (40/14) multilayer also has
long, continuous layer planes but has rougher interfaces, consistent
with the SAXS data.

Transmission electron selected-area diffraction patterns for the
(30/20) and (40/14) films are shown in parts c and d of
Figure~\ref{TEMlowres}.  The (30/20) films shows only a Fe(011) ring,
consistent with the high-angle x-ray diffraction scans.  The (40/14)
film, on the other hand, displays spots corresponding to the (011) and
(002) reflections seen using x-rays.  The presence of spots rather
than rings in the (40/14) image implies the presence of large,
oriented crystallites in the film.  Most interestingly, the (40/14)
image includes a faint spot near what would be the Fe(001) position
were the Fe(001) peak not forbidden by symmetry in the bcc crystal
structure.  The (001) peak is allowed in the B2 and DO$\rm _3$ crystal
structures.  The B2 structure is found in the equilibrium phase
diagram only at 10-22\% Si range of composition,\cite{binary} but
workers at ETH have grown this crystal structure throughout the range
of composition on Si substrates using MBE.\cite{onda} The DO$_3$ phase
found in the equilibrium phase diagram is Fe$_3$Si, which is
ferromagnetic.\cite{binary} Clearly a ferromagnetic spacer phase is
not consistent with the observation of antiferromagnetic interlayer
coupling, although a non-stoichiometric DO$_3$-structure phase might
have different magnetic order.  The B2 and $\epsilon$ iron silicide
phases have both been previously suggested as possible candidates for
the spacer layer in AF-coupled Fe/Si
multilayers.\cite{fullerton,foiles2,dekoster} The position of the
(001) TEM spots is not consistent with the $d$-spacings of the
$\epsilon$ phase.

According to the powder-diffraction files for the B2 and DO$_3$
structures, only the (111) peak of the fcc-family DO$_3$ does not
coincide with a B2 peak.  The (111) peak would be expected to be very
weak in the diffraction patterns formed from cross-sectional specimens
of the film. The reason is that a small number of grains contributes
to the cross-sectional image, and the probability of sampling a grain
with its (111) planes in the observable direction is small because of
the random in-plane orientation.  Future work will include electron
diffraction studies of a (40/14) specimen prepared in the plan-view
geometry, where the number of grains which are sampled is considerably
larger and the odds of observing the fcc (111) peak are improved.

High-resolution TEM images of the (30/20) and (40/14) multilayers are
displayed in Figure~\ref{TEMhires}.  The (30/20) film is shown in
\ref{TEMhires}a to have a crystalline Fe layer and amorphous spacer
layer, similar to the morphology seen before in
Mo/Si\cite{stearns,holloway} and Co/Si multilayers.\cite{miura} The
(40/14) multilayer in Fig.~\ref{TEMhires}b on the other hand is made
up entirely of crystalline layers.  The coherence between the Fe and
silicide spacer is clearly evidenced by the continuity of atomic layer
planes from the Fe layer into the spacer.  Some crystallites in the
(40/14) film extend all the way from the substrate to the surface of
the film.  The small coherence lengths observed in x-ray diffraction
data for the uncoupled thicker-Si films are explained by the presence
of the amorphous layers.  The lack of crystallinity in the spacer
layer of t$\rm _{Si}$ = 20\AA\ films is presumably due to insufficient
time for full interdiffusion and ordering in the thicker layers.  A
kinetic mechanism for the lack of crystallization is supported by
experiments which show that intentional placement of Fe in the Si
layer allows thicker spacer layers to
crystallize.\cite{foiles,mattson}

Another striking feature of the image in Figure~\ref{TEMhires}b is the
periodic modulation that occurs in the silicide spacer layer.  The
modulation originates from scattering by inequivalent planes of atoms.
Simulation of this image using a multiple-scattering computer
calculation may be helpful in positively identifying the crystal
structure of the spacer layer phase.

Dark-field images of the (40/14) multilayer can help answer questions
about the texture of the film as well.  Figure~\ref{TEMdark}a shows
the same bright-field image as in Figure~\ref{TEMlowres}b.  Dark-field
images were formed using (001), (002) and (011) spots from the
diffraction pattern shown in Figure~\ref{TEMlowres}d.  The resulting
micrographs are shown in Figures~\ref{TEMdark}b, c, and d
respectively.  Panels a and b of this figure show the same region of
the (40/14) multilayer.  The brightness of the spacer layers in this
dark-field image demonstrates that the (001) reflection does indeed
come from the spacer layer and is not the forbidden (001) spot of bcc
Fe.  Figures~\ref{TEMdark}c and d also show the same region (although
a different region than panels a and b).  The bright areas in these
two images are the complement of one another; where one is bright, the
other is dark and vice versa.  The dark-field images in panels c and d
of Figure~\ref{TEMdark} demonstrate convincingly that the orientation
of the film evolves from predominantly (011) to predominantly (002) as
the thickness increases.  The reason for the change in orientation
with film thickness is not obvious; it may be related to the
bilayer-period-number dependence discussed in
Section~\ref{bilayersec}.

The effect of varying the Fe thickness has also been studied.
Magnetic properties for films with 20\AA\ $\le$ t$\rm _{Fe}$ $\le$
50\AA\ are found to change only slightly in keeping with the expected
inverse proportionality of the saturation field with t$\rm
_{Fe}$.\cite{demokritov} SAXS peaks tend to broaden and even split
with increasing Fe thickness, indicating increased disorder in the
layering.  The splitting of these peaks may indicate different bilayer
periods in areas of the film with the (011) and (001) textures.  When
the Fe is made less than 20\AA\ thick, the Fe high-angle diffraction
peaks disappear and so does the AF coupling.  The disappearance of
crystalline Fe peaks near t$\rm _{Fe}$ = 20\AA\ is consistent with
previous results on evaporated Fe/Si multilayers.\cite{dufour} Thus
poor crystallinity of the Fe layers appears to suppress the interlayer
coupling even when the Si thickness is favorable.  The lack of AF
coupling in films with poorly crystalline Fe may be related to the
lack of a template for the crystalline iron silicide spacer to grow
on.

\subsection{Dependence of Properties on Growth Temperature and Post-Growth
Annealing}
\label{tempsec}

Depositing the multilayers at different substrate temperatures is an
obvious way of influencing the composition and crystallinity of the
spacer layer phase in the Fe/Si multilayers.  Fullerton has suggested
that the interlayer of Fe/Si multilayers is improved by
high-temperature growth.\cite{fullerton5} We have grown films on glass
substrates at various temperatures between -150 and +200$^{\circ}C$.
The effect of substrate temperature on the interlayer coupling of the
films is illustrated in Figure~\ref{magtemp}, where magnetization
curves for three (Fe40\AA/Si14\AA)x40 multilayers grown at
-150$^{\circ}$C, +60$^{\circ}$C (nominal RT) and +200$^{\circ}$C are
shown.  The data show that as the substrate temperature increases the
saturation field increases indicating larger AF coupling.  The
saturation magnetization also decreases, suggesting a larger degree of
interdiffusion in the films grown at higher temperatures.

The suspicion that more interdiffusion occurs at higher substrate
temperatures is confirmed by examination of the SAXS spectra for the
three films, shown in Figure~\ref{lowxrdtemp}.  The film grown at
reduced temperature has 7 peaks while the film grown at nominal RT has
5 and the film grown at +200$^{\circ}$C has only 4.  Quantitative
modelling of low-angle x-ray data has shown that the suppression of
higher-order peaks may be due to either interdiffusion or cumulative
roughness.\cite{fullerton3,payne} Certainly larger cumulative
roughness could also occur at higher growth temperatures, but one
would expect very rough growth to suppress AF coupling due to an
increased number of pinholes and larger magnetostatic interlayer
coupling.\cite{altbir} Since higher growth temperatures seem to
enhance rather than suppress the coupling, it seems more likely that
high substrate temperatures are promoting interdiffusion rather than
roughness.  Studies of Mo/Si multilayers showed that a growth
temperature of 150$^{\circ}$C gives maximum SAXS reflectivity, which
the authors attribute to greater interface smoothness than for RT
deposition.\cite{stock} Smaller bilayer periods in multilayers grown
at higher temperatures support the claim of increased interdiffusion.
Fitting Eqn.~\ref{lowbragg} to peak positions from
Figure~\ref{lowxrdtemp} gives $\Lambda$ = 52.7, 49.3, and 43.8 \AA\
respectively for the -150$^{\circ}$, +60$^{\circ}$, and +200$^{\circ}$
multilayers versus the nominal value of 54\AA.

Higher substrate temperatures may also promote ordering of the Fe and
Si atoms in the crystalline spacer layer.  In the fully ordered B2
phase, the Fe and Si atoms sit on different simple cubic sublattices.
The sublattice order can occur irrespective of whether or not the Fe
to Si ratio is 1:1.  It is interesting to speculate whether the AF
coupling is dependent on the degree of ordering in the spacer layer.
An ordering-dependent coupling seems plausible in light of the
Fermi-surface theories of coupling in metal/metal
multilayers.\cite{bruno2,stiles} A well-ordered B2 or DO$\rm _3$ phase
would have more well-defined Fermi surface features than a random
solid solution.  Unfortunately the (001) silicide peak has only been
observed by TEM, making experimental attempts to address this issue
difficult.  Further studies with x-ray diffraction and soft x-ray
fluorescence are underway.

The crystallinity of the films also varies with growth temperature.
Surprisingly, films grown at both low and high temperatures on glass
substrates always have only the (011) texture, while films grown at
nominal RT often have mixed (001) and (011) textures.  The multilayers
deposited on heated and cooled substrates do differ greatly in that
those grown at low temperature have amorphous spacer layers, while
those grown at high temperatures have long crystalline coherence
lengths.  The reasons for the strange temperature dependence of growth
texture are not understood, although one presumes that they have to do
with the kinetics of growth.  It is not clear why the (001) texture
should appear at all, although it has also been seen in Mo/Ge
multilayers.\cite{wilson} An oscillatory dependence of film texture on
spacer layer thickness and deposition conditions has been reported for
NiFe/Cu multilayers grown by IBS.\cite{nakatani} The (001) texture has
not been reported in polycrystalline magnetron-sputtered Fe/Si
multilayers, and may be due to some peculiarity of IBS growth.


A logical extension to the growth temperature studies is to try
annealing the Fe/Si multilayers grown at lower substrate temperatures
to see if their properties evolve towards those of the multilayers
grown at higher temperatures.  As far as the magnetic properties are
concerned, the answer is ``no.''  Annealing the uncoupled RT-grown
(Fe30\AA/Si20\AA)x40 and low-temperature-grown (Fe40\AA/Si14\AA)x40
multilayers at +200$^{\circ}$C for two hours had almost no effect on
their magnetic properties beyond a slight magnetic moment reduction.
A subsequent 300$^{\circ}$C anneal for two hours once more produced a
moment reduction and a decrease in coercive field in the uncoupled
multilayers.  A very low coercive field for annealed Fe/Si films is
not surprising given the well-known softness of Fe-Si alloys.  A
300$^{\circ}$C anneal even eliminated the interlayer exchange coupling
of a RT-grown (Fe40\AA/Si14\AA)x50 film used as a control.  For this
(40/14) multilayer, the 300$^{\circ}$C anneal caused the SAXS peaks to
narrow and reduced their number from 5 to 4.  At the same time the
bilayer period decreased from 49.4\AA\ to 46.0\AA.  High-angle x-rays
spectra (not shown) indicated that the Fe lattice constant slightly
decreased, which is consistent with increased diffusion of Si in the
Fe layer.\cite{foiles2} These x-ray and magnetization results imply
that annealing primarily promotes interdiffusion of the Fe and
silicide layers.  With sufficient interdiffusion the spacer layer may
become ferromagnetic, which would explain the suppression of
antiferromagnetic interlayer coupling.  These Fe/Si multilayers show
less thermal stability than Mo/Si multilayers with comparable layer
thicknesses, which do not show changes in SAXS spectra until
400$^{\circ}$C.\cite{stock} There was no sign of the solid-state
amorphization previously observed in Fe/Si multilayers with thicker
layers.\cite{gupta}

Whatever process occurs during annealing, it does not enhance the
interlayer coupling the way that +200$^{\circ}$C growth does.  This is
hardly surprising given that annealing will tend to drive the
multilayer towards its equilibrium state, presumably a mixture of
different iron silicide phases.  There is no reason to think that the
crystalline Fe/Fe$_x$Si$_{1-x}$ multilayer should be an
intermediate phase during the annealing.  In the future the kinetics
of Fe/Si multilayer growth at different substrate temperatures will be
investigated further by employing an ion-assist gun to improve atomic
surface mobility.

\subsection{Dependence of Properties on Number of Bilayers}
\label{bilayersec}

One puzzling aspect of the interlayer exchange coupling in the Fe/Si
system has been the dependence of its strength on the number of
bilayers in the multilayer.  This trend is illustrated in
Figure~\ref{binumber}, where magnetization curves for
(Fe40\AA/Si14\AA)xN multilayers with 2, 12 and 25 repeats are
displayed.  (The 2-repeat multilayer is just an Fe/Si/Fe trilayer.)
Although the trilayer has magnetic properties like bulk Fe, the
25-repeat multilayer data has a magnetization curve similar to the
40-repeat multilayer data shown above.  The magnetization curve for
the 12-repeat multilayer falls in between that for the thicker and
thinner films.  Evidence for AF coupling which is stronger near the
top of an Fe/Si multilayer than near the substrate has previously been
described by Fullerton {\it et al.}\cite{fullerton4} Presumably the
increase of coupling with bilayer-number is a manifestation of the
same phenomenon.  The interlayer coupling in Co/Cu multilayers also
increases with the number of bilayer periods up to about 25
bilayers.\cite{rupp}

One would not expect interlayer coupling that is quantum-mechanical in
nature to be affected much by total film thickness.  The unusual
thickness dependence therefore raises the question of whether there is
quantum-mechanical coupling at all, or whether some other mechanism
might determine the shape of the magnetization curves.  Disordered
magnetic materials such as small amorphous Fe particles can have low
remanence and high saturation fields without any layering at all. The
magnetization curves of these Fe particles are in fact quite similar
to those of the Fe/Si multilayers.\cite{grinstaff} This resemblance
might lead to speculation that the topmost Fe layers in Fe/Si
multilayers are discontinuous and that the magnetic properties are
dominated by particle shape.  However, the existence of half-order
peaks in polarized neutron reflectometry measurements in the IBS-grown
Fe/Si multilayers\cite{ankner} and the magnetron-sputtered
multilayers\cite{fullerton4} gives unambiguous evidence that the
magnetic properties are due to magnetic order rather than structural
disorder.  In addition, TEM pictures such as Figure~\ref{TEMlowres}
show that the Fe layers are continuous in films with both high and low
saturation fields.

How then does the number of bilayer periods influence the AF coupling
strength?  It has been suggested that the difference between thin and
thick multilayers grown at nominal RT is that the substrates of thick
multilayers have time to rise to a higher temperature (about
+60$^{\circ}$C for our system) during the longer
growth.\cite{fullerton5} This idea seems reasonable in light of the
larger coupling in samples grown on heated substrates as described
above.  In order to investigate this idea, a
(Fe100\AA/Si14\AA/Fe100\AA) film was grown on glass at
+200$^{\circ}$C.  The magnetization curve for this film is shown in
Figure~\ref{trilayers}.  Also shown in this figure are data for a
(Fe100\AA/Si14\AA/Fe100\AA) trilayer deposited at nominal RT and for a
(Fe100\AA/Si14\AA/Fe100\AA) trilayer deposited at +200$^{\circ}$C,
both grown on a 500\AA-thick a-Si buffer.  The trilayer deposited
directly on glass at elevated temperature has only slightly less
remanence and higher saturation field than the trilayer grown at RT
whose data are shown in Figure~\ref{binumber}.  This result implies
that it is not substrate temperature alone which causes bilayer-number
effects.  The magnetization curves of the trilayers grown on buffer
layers, on the other hand, look much more like typical t$\rm _{Fe}$ =
40\AA\ 40-repeat multilayer results.  An epitaxial
(Fe100\AA/Si14\AA/Fe100\AA) trilayer grown directly on an MgO(001) at
+200$^{\circ}$C substrate also has strong AF coupling (data not
shown).  Undoubtedly the strong AF coupling of the trilayer grown
directly on the MgO is due the superior surface quality of the
single-crystal substrate.

The take-away lesson from all of these results is that substrate
roughness is probably responsible for the reduced interlayer coupling
in (Fe40\AA/Si14\AA) multilayers with a low number of bilayers.
Conformal growth may propagate this roughness up from the substrate
into the multilayer.  Parkin {et al.} have found that the interlayer
coupling in MBE-grown Co/Cu multilayers is very sensitive to the
substrate and the buffer layer type, perhaps due to pinholes through
the Cu layers.\cite{parkin} Presumably thin Fe layers grown directly
on glass are so wavy that pinhole and magnetostatic coupling dominate
the interlayer interactions for the first few bilayer periods.  Recent
calculations show that magnetostatic effects associated with
propagating roughness can give interlayer ferromagnetic coupling of
the same order of magnitude as the coupling derived from quantum-well
effects.\cite{altbir} Ongoing polarized neutron reflectivity
experiments may give more information on the variation of the coupling
with position in the thicker multilayers.\cite{ankner}

\subsection{Growth on Single-Crystal Substrates}

That Fe films can be grown epitaxially on MgO and $\rm Al_2O_3$
substrates is well-known.\cite{metoki} One might therefore expect to
be able to grow high-quality Fe/Si superlattices on these substrates.
Figure~\ref{epihigh}a) shows high-angle x-ray diffraction spectra for
a purely (001)-oriented (Fe40\AA/Si\AA)x60 multilayer grown on
MgO(001).  The spectrum in Figure~\ref{epihigh}b) is data for a highly
(011)-oriented (Fe40\AA/Si14)x46 multilayer grown on $\rm Al_2O_3$.
Both multilayers were deposited at +200$^{\circ}$C.
Figure~\ref{epihigh}c) shows a $\phi$ scan for the MgO (110) and Fe
(110) peaks for the film on the MgO substrate.  These sets of peaks
are offset from one another by 45$^{\circ}$ in $\phi$, confirming the
well-known epitaxial relation Fe(001) $\|$ MgO(001) and Fe(110) $\|$
MgO(100).\cite{metoki} The $\phi$ scans for the $\rm Al_2O_3$
substrate show that this film is only weakly oriented in-plane.
Mattson {et al.} have previously grown Fe/FeSi multilayers on $\rm
Al_2O_3$, but they did not comment on the orientation of the
multilayer.\cite{mattson2} Rocking curves widths for both films are
about 1$^{\circ}$ wide, indicating a considerably smaller mosaic than
for the multilayers grown on glass.  SAXS data for the multilayers on
single-crystal substrates are comparable to the data for films grown
on glass.

The films grown on MgO are the only purely (001)-textured Fe/Si
multilayers produced by IBS so far.  Dekoster {et al.} have grown
epitaxial Fe/FeSi multilayers on MgO(001) by MBE, but they do not
present any x-ray diffraction data or magnetization
curves.\cite{dekoster} Magnetization curves of films grown on
single-crystal substrates (not shown) are qualitatively similar to
those grown on glass or oxidized Si substrates.  The only differences
are that the saturation fields are higher for the epitaxial samples
and that magnetocrystalline anisotropy effects are observed.  The
magnetocrystalline anisotropy energies of epitaxial trilayers grown on
MgO and Ge are similar to bulk Fe.\cite{michel}

The shape of the high-angle peaks plus superlattice satellites are
described by a theory due to Fullerton {\it et al.}\cite{fullerton3}
Application of this theory to the Fe/Si multilayers is difficult
because the silicide lattice constant, the thickness of the remaining
pure Fe and the thickness of the silicide spacer can be estimated only
roughly.   A precise determination of the silicide lattice constant
should make a quantitative analysis of these satellite features possible.

\section{Discussion}

Fe and Si appear to be the only known transition-metal/semiconductor
combination in which the two elements interdiffuse to form a
crystalline spacer layer with coherent interfaces.  The reasons why
this unusual morphology occurs in the Fe/Si system are unknown but
likely involve a high rate of Fe diffusion into a-Si and a low heat of
crystallization of the iron silicide compound.  A detailed discussion
of these issues is beyond the scope of this paper.

Three different crystal structures have been proposed for the
crystalline spacer layer of the Fe/Si multilayers.  The $\epsilon$
phase can be eliminated on the basis of the electron diffraction
patterns and TEM dark field images presented here.  The B2 and DO$_3$
crystal structures are better lattice-matched to Fe than
$\epsilon$-FeSi or $\alpha$- and $\beta$-FeSi$_2$.  The lattice
constant of the B2 phase was reported by M\"ader and coworkers to be
2.77\AA, only 3.1\% different from Fe.\cite{mader} The lattice
constant of the $\epsilon$ phase is 4.46\AA,\cite{mader} which matches
the Fe(110) plane only in the energetically unfavorable (210)
direction.\cite{mattson}


Recent conversion-electron M\"ossbauer data are interpreted in
support of the B2 crystal structure, although the possibility of the
DO$_3$ phase was not considered in that study.\cite{dekoster} It is
plausible that the B2 or DO$_3$ structures form in rapid,
far-from-equilibrium growth conditions because of their small unit
cells.  Since silicon deposited at low substrate temperatures is
amorphous, the most likely scenario is the following.  Silicon
deposited on a crystalline Fe layer goes down amorphous and diffuses
only slightly into the Fe.  Subsequently deposited Fe atoms diffuse
rapidly into the amorphous Si, analogous to what happens during the
growth of Mo/Si multilayers.\cite{stearns,holloway} During the
diffusion of Fe into Si, crystallization of the silicide occurs,
possibly driven by the heat of mixing or by the kinetic energy of the
incident Fe atoms.  Growth of the crystalline phase may proceed upward
from the lattice-matched Fe template, or downward from the atomically
bombarded film surface.  If the growth of the crystalline silicide
phase proceeds downward from the film surface, one might expect to
see some crystalline silicide in the high-resolution TEM image for the
t$\rm _{Si}$ = 20\AA\ film (Figure~\ref{TEMhires}b).  The lack of any
evidence for crystalline silicide in this image suggest that the
crystallization proceeds upward from the iron/silicide interface, not
downward from the film surface.

It is difficult to determine how realistic this model for growth of
the crystalline silicide is since the Fe/Fe-Si and Si-Fe/Fe interfaces
appear identical in Figure~\ref{TEMhires}b.  In contrast, the Mo/Si
and Si/Mo interfaces in Mo/Si multilayers appear quite different from
one another.\cite{stearns,holloway} In the Mo/Si multilayers, an
amorphous MoSi$_2$ region appears which is thicker at the Mo/Si
interface than at the Si/Mo interface.  Detailed TEM studies of
multilayers with t$\rm _{Si}$ larger than 20\AA\ may help to answer
whether amorphous silicides can occur in IBS-grown Fe/Si multilayers.

Using the B2 phase lattice constant reported by the Z\"urich
group,\cite{mader} we can estimate the expected bilayer period of a
nominal Fe/Si multilayer in which Fe atoms diffuse into the Si layer
up to a 1:1 stoichiometry.  The spacing between the Fe and Fe$\rm
_x$Si$\rm _{1-x}$ layers is taken as the average of the interplanar
spacings of the two materials.  The result of this rough calculation
is that an (Fe40\AA/Si14\AA) multilayer which interdiffuses up to the
1:1 stoichiometry should form a (Fe33.2\AA/FeSi16.3\AA) multilayer
with a bilayer period of 49.4\AA.  The missing bilayer period
predicted from this model is 4.6\AA, in the middle of values on the
x-axis of Figure~\ref{missmom}.  One can also calculate the expected
magnetic moment reduction assuming that Fe atoms in the silicide layer
have no moment and those in the Fe layer have their full moment.
Under this assumption a calculation predicts 8.2\AA\ of missing Fe
moment, slightly lower than indicated in Figure~\ref{missmom}.  This
calculation neglects the possibility that some Fe atoms in the Fe
layer with Si near neighbors may have reduced magnetic moments.

In the discussion above the possibility has not been mentioned that
the missing bilayer period and magnetic moment are due to an
inaccurate thickness calibration.  This explanation is contradicted by
magnetization and x-ray diffraction measurements on Fe/Ge multilayers,
where measured magnetic moments and bilayer periods are in much closer
agreement with nominal values than for Fe/Si.\cite{michel} The
improved agreement in the case of Fe/Ge multilayers suggests that
interdiffusion is less important in multilayers with Ge spacer layers
than in multilayers with Si spacers.

The main point is that the formation of the B2 silicide does
qualitatively explain the bilayer period reduction observed in the
Fe/Si multilayers.  The underlying reason for the bilayer period
reduction is that the silicide which forms is denser than both Fe and
Si.  This situation is similar to that observed in other metal/Si
multilayers\cite{fullerton2,holloway} except that in the other
multilayers the silicide remains amorphous.

Confirmation that the spacer layer phase has the B2 or DO$\rm _3$
structure is important for understanding the coupling mechanism in
these compounds.  Both the B2 and DO$\rm _3$ phases are known to be
metallic for some ranges of composition.\cite{vonkanel,kudrnovsky}
Thus the present results and those of other
workers\cite{fullerton,dekoster} suggest that Fe/Si is really a
metal/metal multilayer.  The origin of the interlayer coupling is then
likely to be described by the same theories as describe coupling in
Co/Cu and Fe/Cr multilayers.\cite{bruno2,stiles} Fe/Si multilayers may
therefore not be a good test case for theories which model interlayer
exchange coupling across insulators.\cite{bruno,singh}

In the discussion above the possibility has been neglected that the
amorphous spacer layer in the thick-Si films may also be metallic.  If
both the thick amorphous spacers and the thin crystalline spacers are
metallic silicides, then it must be the crystallinity that is the
essential feature for the existence of AF interlayer coupling.  Up to
now there have been no reports of AF coupling across amorphous
metallic spacer layers.  Toscano {\rm et al.} have reported AF
coupling across amorphous silicon spacer layers.\cite{toscano} These
Fe/a-Si/Fe trilayers were prepared at low temperature so as to
suppress interdiffusion.\cite{toscano} The character of AF coupling in
the a-Si spacer trilayers is likely quite different than in the
multilayers described in this study, where substrate heating increases
the strength of coupling.

At the moment there is no direct evidence regarding the metallic or
insulating nature of the amorphous spacer layers found in the
(Fe30\AA/Si20\AA) multilayers.  Temperature-dependent current-in-plane
resistivity measurements suggest that both crystalline and amorphous
spacer layers in Fe/Si multilayers are poorly conducting.\cite{michel}
Fe$_{70}$Si$_{30}$ and Fe$_{65}$Si$_{35}$ amorphous alloys have a
temperature-independent resistivity, suggesting non-metallic
behavior.\cite{marchal} Overall the evidence suggests that the
amorphous spacer layers in (Fe30\AA/Si20\AA) multilayers are not
metallic, but spectroscopic measurements like soft x-ray
fluorescence\cite{carlisle} are needed for confirmation.  The
interesting question as to whether there can be AF interlayer coupling
across an amorphous metal spacer layer must then be left for another
study.

\section{Conclusions}

An extensive study of the growth of Fe/Si multilayers by ion-beam
sputtering has been performed.  The crystalline quality of the films
is better when they are grown with thick Fe layers, with thin Si
layers, at high temperature, and on single-crystal substrates.
Improved growth conditions lead to higher saturation fields and lower
remanence in magnetization curves.  Measured bilayer periods are
consistently shorter in these multilayers than the nominal value,
suggesting formation of a dense silicide phase in the spacer layer.
Despite considerable interdiffusion in the multilayers, a strong
composition modulation along the growth direction is maintained as
evidenced by SAXS measurements.

There are two surprising results from this study.  One is that the
films grow on glass with a mixed (011) and (001) texture near nominal
RT and with a pure (011) texture at higher and lower temperatures.
The other surprise is that the strength of the interlayer coupling
depends strongly on the number of bilayer periods in films with thin
Fe layers.  This latter result is explained on the basis of substrate
surface roughness.

Unraveling the behavior of the Fe/Si multilayer system has proven to
be a considerably more complex task than understanding the Fe/Cr or
Co/Cu multilayer systems.  The reason is that compound formation at
the Fe/Si interface is crucial to understanding the AF interlayer
coupling.  Identification of possibly disordered phases in the spacer
layer of a multilayer continues to be an experimental challenge.
Mounting evidence suggests that the spacer layer in the AF-coupled
Fe/Si multilayers is metallic and crystalline and that the Fe/Si
interlayer coupling therefore has the same origin as in metal/metal
multilayers.

\bigskip

\paragraph*{Acknowledgements}

\noindent We would like to thank P.E.A. Turchi, T.W. Barbee Jr., T.P. Weihs,
E.E. Fullerton, Y. Huai and E.C. Honea for helpful discussions, and
B.H. O'Dell and S. Torres for technical assistance.  Further thanks go
to C.-T. Wang of Stanford for the four-circle x-ray diffractometry and to
Sandia National Lab for use of their electron microscope for HREM
work.  Part of this work was performed under the auspices of the
U.S. Department of Energy by LLNL under contract No. W-7405-ENG-48.

\clearpage


\clearpage

\begin{figure}
\caption{Schematic plan of the ion-beam sputtering system.}
\label{chamber}
\end{figure}

\begin{figure}
\caption{Magnetization curves for (Fe30\AA/Si20\AA)x50 and
(Fe30\AA/Si14\AA)x50 multilayers grown on glass substrates at nominal
RT during the same deposition run.  Plotted on the y-axis is the
observed magnetization of the films divided through by the calculated
magnetization of an equivalent thickness of bulk Fe.  The
(Fe30\AA/Si20\AA)x50 multilayer has soft magnetic properties much like
bulk Fe, while the (Fe30\AA/Si14\AA)x50 multilayer exhibits AF
interlayer coupling.}
\label{sithick}
\end{figure}

\begin{figure}
\caption{X-ray diffraction spectra at small-angle for the same films
whose magnetization curves are shown above.  Broader peaks show that
there is more disorder in layering for the AF-coupled film with t$\rm
_{Si}$ = 14\AA.  Using Equation~1, these data give bilayer periods
$\Lambda$ = 41.82\AA\ for the nominal (Fe30\AA/20\AA)x50 film and
$\Lambda$ = 38.10\AA\ for the nominal (Fe30\AA/Si14\AA)x50 film.}
\label{sithickxrdlow}
\end{figure}

\begin{figure}
\caption{Missing Fe magnetic moment expressed as an equivalent thickness of
Fe plotted versus missing bilayer period as obtained from fits to
small-angle x-ray diffraction data.  Symbols indicate different
nominal Si layer thicknesses and different film textures.  The film
labelled ``LN'' was grown on a LN-cooled substrate; all others were
grown at nominal RT.  All multilayers have 40 or 50 repeats and were
grown on either glass or oxidized Si substrates.}
\label{missmom}
\end{figure}

\begin{figure}
\caption{High-angle spectra for two Fe/Si multilayers showing the Fe
(011) and (002) peaks.  The t$\rm _{Si}$ = 20\AA\ film is
predominantly (011)-textured, while the AF-coupled film with t$\rm
_{Si}$ = 14 \AA\ has mixed (011) and (001) textures.  No x-ray
diffraction peaks which could be indexed to crystalline silicon or
silicide spacer layer phases have been observed in any Fe/Si
multilayer.  A superlattice satellite just below the Fe(002) peak is
labelled ``-1.''}
\label{sithickxrdhi}
\end{figure}

\begin{figure}
\caption{Cross-sectional TEM images (a and b) and selected area
diffraction patterns (c and d) for the same (Fe30\AA/Si20\AA)x50
multilayer and a (Fe40\AA/Si14\AA)x50 multilayer grown that shows
strong AF coupling.  a) and b) show that the Fe/Si multilayers have
layers which are continuous for large lateral distances.  There is no
sign of propagating roughness or columnar growth.  c) The (30/20)
multilayer shows only an Fe(011) ring.  d) The (40/14) film shows
(011) and (002) spots plus a faint spot at the (001) position
(indicated by an arrow).}
\label{TEMlowres}
\end{figure}

\begin{figure}
\caption{High-resolution TEM images of the same films whose
low-resolution images are shown above.  a) (Fe30\AA/Si20\AA)x50
multilayer image showing amorphous silicide layers between
polycrystalline Fe layers.  b) (Fe40\AA/Si14\AA)x50 multilayer image
showing crystalline coherence between the polycrystalline Fe layers
and iron silicide spacer layers.  There is no amorphous layer
present.}
\label{TEMhires}
\end{figure}

\begin{figure}
\caption{a) The same bright field TEM micrograph of the (Fe40\AA/Si14\AA)x50
multilayer as is shown in Figure 6b.  b) A dark-field image of the
same region of the (40/14) multilayer.  This dark-field image was
formed using the (001) reflection.  Comparison with the bright field
image shows that the (001) reflection originates from the Si substrate
and the spacer layers.  c) and d) Dark-field images formed from (002)
and (011) reflections.  Image c) shows that planes with (002)
orientation predominate near the film surface.  Image d) shows that
planes with (011) orientation predominate near the substrate.   The
film surface is on the top of all these images.}
\label{TEMdark}
\end{figure}

\begin{figure}
\caption{Magnetization curves for three (Fe40\AA/Si14\AA)x40
multilayers grown on glass substrates at -150$^{\circ}$,
+60$^{\circ}$C and +200$^{\circ}$C.  The increase of the saturation
field with increasing substrate temperature indicates an increase in AF
coupling.  Note that the saturation magnetization also decreases
slightly with increasing substrate temperature.}
\label{magtemp}
\end{figure}

\begin{figure}
\caption{Small-angle x-ray diffraction spectra for three
(Fe40\AA/Si14\AA)x40 multilayers grown on glass substrates at
-150$^{\circ}$C, +60$^{\circ}$C and +200$^{\circ}$C.  The
disappearance of higher-order peaks at higher substrate temperatures
is an indication of greater interdiffusion.}
\label{lowxrdtemp}
\end{figure}

\begin{figure}
\caption{Magnetization curves for 2-, 12- and 25-repeat
(Fe40\AA/Si14\AA) multilayers grown during the same deposition run at
nominal RT on glass substrates.  The 2-repeat multilayer (really an
Fe/Si/Fe trilayer) shows no signs of AF coupling.  The 12-repeat
multilayer appears to have a smaller coupling than the 25-repeat one.}
\label{binumber}
\end{figure}

\begin{figure}
\caption{Magnetization curves of three (Fe/Si/Fe) trilayers.  The
open circles are data for an (Fe100\AA/Si14\AA/Fe100\AA) film grown
directly on glass at +200$^{\circ}$C.  The filled circles are data on a
(Fe100\AA/Si14\AA/Fe100\AA) film grown at +200$^{\circ}$C on a 500\AA\
a-Si buffer layer on glass.  The solid curve is for a
(Fe100\AA/Si14\AA/Fe100\AA) film grown at nominal RT on a 500\AA\ a-Si
buffer layer on glass.  The coupling is stronger in the film grown at
high temperature on a buffer than in either of the other two films.}
\label{trilayers}
\end{figure}

\begin{figure}
\caption{High-angle x-ray diffraction spectra from Fe/Si multilayers
grown on single-crystal substrates.  a) Data for a
(Fe40\AA/Si14\AA)x60 multilayer grown on MgO(001).  The Fe(002) peak
is shown with 5 satellites centered at 64.77$^{\circ}$.  b) Data for a
(Fe40\AA/Si14\AA)x46 multilayer grown on $\rm Al_2O_3$(0\=211).
Visible in the spectrum are the $\rm Al_2O_3$ (0\=211) peak at
37.79$^{\circ}$ and the Fe(011) peak centered at 44.99$^{\circ}$ with
its 4 satellites.  c) $\phi$ scans plotted on a logarithmic scale for
the MgO and Fe (110) peaks of the (Fe40\AA/Si14\AA)x60 multilayer
grown on MgO. The Fe (100) direction is parallel to the MgO (110), as
expected, but a small amount of material with a secondary orientation
is also visible.}
\label{epihigh}
\end{figure}

\end{document}